\documentclass[aps,prl,reprint,twocolumn,showpacs,superscriptaddress,amsmath,amssymb]{revtex4-1}

\usepackage{graphicx}
\usepackage{dcolumn}
\usepackage{bm}
\usepackage{color}

\usepackage[latin1]{inputenc}
\usepackage{amsfonts}
\usepackage{amsmath}
\usepackage{amssymb}
\usepackage{amsthm}
\usepackage{bbold}
\usepackage[american]{babel}
\usepackage[T1]{fontenc}
\usepackage{xspace}
\usepackage{bbm}
\usepackage[all]{xy}

\begin{document}


\title{Liquid methane at extreme temperature and pressure: Implications for models of Uranus and Neptune}

\author{Dorothee Richters}
\affiliation{%
Institute of Mathematics and 
Center for Computational Sciences, Johannes Gutenberg University Mainz, Staudinger Weg 9, D-55128 Mainz, Germany
}%

\author{Thomas D. K\"uhne}%
\email{kuehne@uni-mainz.de}
\affiliation{%
Institute of Physical Chemistry and 
Center for Computational Sciences, Johannes Gutenberg University Mainz, Staudinger Weg 7, D-55128 Mainz, Germany
}%

\date{\today}

\begin{abstract}
We present large scale electronic structure based molecular dynamics simulations of liquid methane at planetary conditions. In particular, we address the controversy of whether or not the interior of Uranus and Neptune consists of diamond. In our simulations we find no evidence for the formation of diamond, but rather sp$^2$-bonded polymeric carbon. Furthermore, we predict that at high temperature hydrogen may exist in its monoatomic and metallic state. The implications of our finding for the planetary models of Uranus and Neptune are in detail discussed.
\end{abstract}

\pacs{31.15.aq, 71.15.Pd, 96.15.Nd, 96.15.Pf}
\keywords{Linear Scaling, Tight-Binding, Car-Parrinello, Molecular Dynamics, Uranus, Neptune, Metallic Hydrogen, Methane}
\maketitle

Being the most abundant organic molecule in the universe, liquid CH$_4$ at high temperature and pressure is of great relevance for planetary science. The here considered pressure and temperature conditions follow the isentrope in the middle ice layers of Neptune and Uranus at a depth of one-third the planetary radius below the atmosphere. The gravity fields and mean densities of the outer gas giants Neptune and Uranus allude to a three-layer model: a relatively small central rocky core composed of iron, oxygen, magnesium and silicon, followed by an ice mantle and a predominantly hydrogen atmosphere. 
The middle ice layer consists of CH$_4$, NH$_3$ as well as H$_2$O and, in spite of its name, is not solid but gaseous in the outer atmospheres and a hot liquid in the interior. At variance to the planetary models of Saturn and Jupiter, the observed values for mass and radius indicate that hydrogen cannot be an integral part of either Neptune and Uranus. Since it is moreover not primordial, the detected abundance of hydrogen in the atmospheres of both planets implies that it may initially originate from deep within the planets and brought to the outmost layer by convection, where it does not substantially contribute to the total mass \cite{hubbard}.

In any case, information on the interior structure of Neptune and Uranus are scarce and experimentally only indirectly accessible by means of Voyager~II flyby measurements \cite{nellisUranus, hubbardNeptune}, shock-wave compression \cite{nellis1, nellis2}, as well as laser-heated diamond anvil cell experiments \cite{benedetti}. Even though CH$_4$ is the most stable hydrocarbon at ambient conditions, based on these shock-wave experiments as well as theoretical ground state calculations \cite{ree1}, it has been suggested that CH$_4$ may dissociate around $P=20$~GPa and $T=2000$~K into H$_2$ and diamond \cite{ross}. While there is little doubt that in the cores of Uranus and Neptune CH$_4$ dissociates into diamond, this would be anyhow rather consequential as it implies that in the interiors of these giant planets there is no CH$_4$ at all, but a huge diamond mine instead.

On the other hand, \textit{ab-initio} molecular dynamics (AIMD) simulations predicted that the formation of diamond is preempted by the appearance of hydrocarbons \cite{ancil}. Notwithstanding that their finding had been subsequently confirmed by laser-heated diamond anvil cell experiments, which, at pressure $P=19$~GPa and temperature $T=2000-3000$~K, indicate the presence of both polymeric carbon as well as diamond \cite{benedetti}. This view was further strengthened by subsequent AIMD simulations, even though none of them found any evidence for diamond formation \cite{kress1, kress2, spanu}. 
Nevertheless, AIMD simulations are particularly appropriate to directly probe CH$_4$ under the extreme pressure and temperature conditions predominating in the middle ice layer, in particular as here in either case covalent bonds are broken and formed. 
Moreover, all of the AIMD simulations show that the intricate interplay between temperature and pressure is essential to grasp CH$_4$ at planetary conditions, where covalent C-H bonds are broken by heat, while compression favors condensation of the dissociated carbon atoms. It is therefore suggestive that in AIMD simulations at even higher pressure, but still in the middle ice layer, carbon may nonetheless spontaneously transform into diamond. However, what causes the large discrepancy in the pressure between theory and experiment when diamond is formed is unknown. 

In this Letter we revisit the behavior of liquid CH$_4$ by means of a novel field-theoretic approach to linear scaling AIMD simulations \cite{ceri1, ceri2, richMethod}. 
However, contrary to previous AIMD simulations, where the considered systems sizes have been rather small \cite{ancil, kress1, spanu}, here we use as many as 1000 CH$_4$ molecules in a periodic cubic simulation box of length $L = 25{.}55\,$\r{A} as our unit cell. All of our calculations have been performed in the canonical ensemble at $T = 2000 - 6000$~K and volume $V=10.04$~cm$^3$/mol, which corresponds to the second-shock at $P = 92$~GPa and $T=4000$~K of a two-stage light-gun shock compression experiment \cite{nellis1}. 

For all of our large scale MD simulations we have employed the self-consistent tight-binding model for hydrocarbons \cite{horsf} as implemented in the CMPTool code \cite{cmpt}. The atoms are propagated by a modified Langevin equation using a discretized time step of $\Delta t = 0.5$~fs \cite{kueh1, kuehne1}. Since we are dealing with a large disordered system at finite temperature, the Brillouin zone is sampled at the $\Gamma$-point only. 

Well-equilibrated and long trajectories are essential to ensure an accurate sampling. To that extend we have at first carefully equilibrated each of our our simulations at $T = 2000, 4000$, and 6000~K, before accumulating statistics for overall 50~ps. Even though dissociation processes typically happen on rather short timescales, it is important to note that the temperature for dissociation and dehydrogenation as determined by direct MD simulations only represents an upper bound. 
A major advantage of our novel grand-canonical simulation technique is that, at variance to conventional ground-state AIMD simulations, excited electrons can be employed. Due to the fact that they are known to dramatically weaken covalent bonds \cite{silvestrelli1, *silvestrelli2}, and therefore may facilitate the dissociation of CH$_4$, we have hence chosen $\beta/k_B$ to be identical with the nuclear temperature. Nuclear quantum effects, such as zero-point energies, are less important for the high temperature regime examined here and are therefore neglected. However, entropy effects have been shown to be very relevant, so that the dissociation of CH$_4$ is supposedly much more sensitive to temperature than it is to pressure \cite{spanu}. 

\begin{figure}
\includegraphics[width=0.475\textwidth]{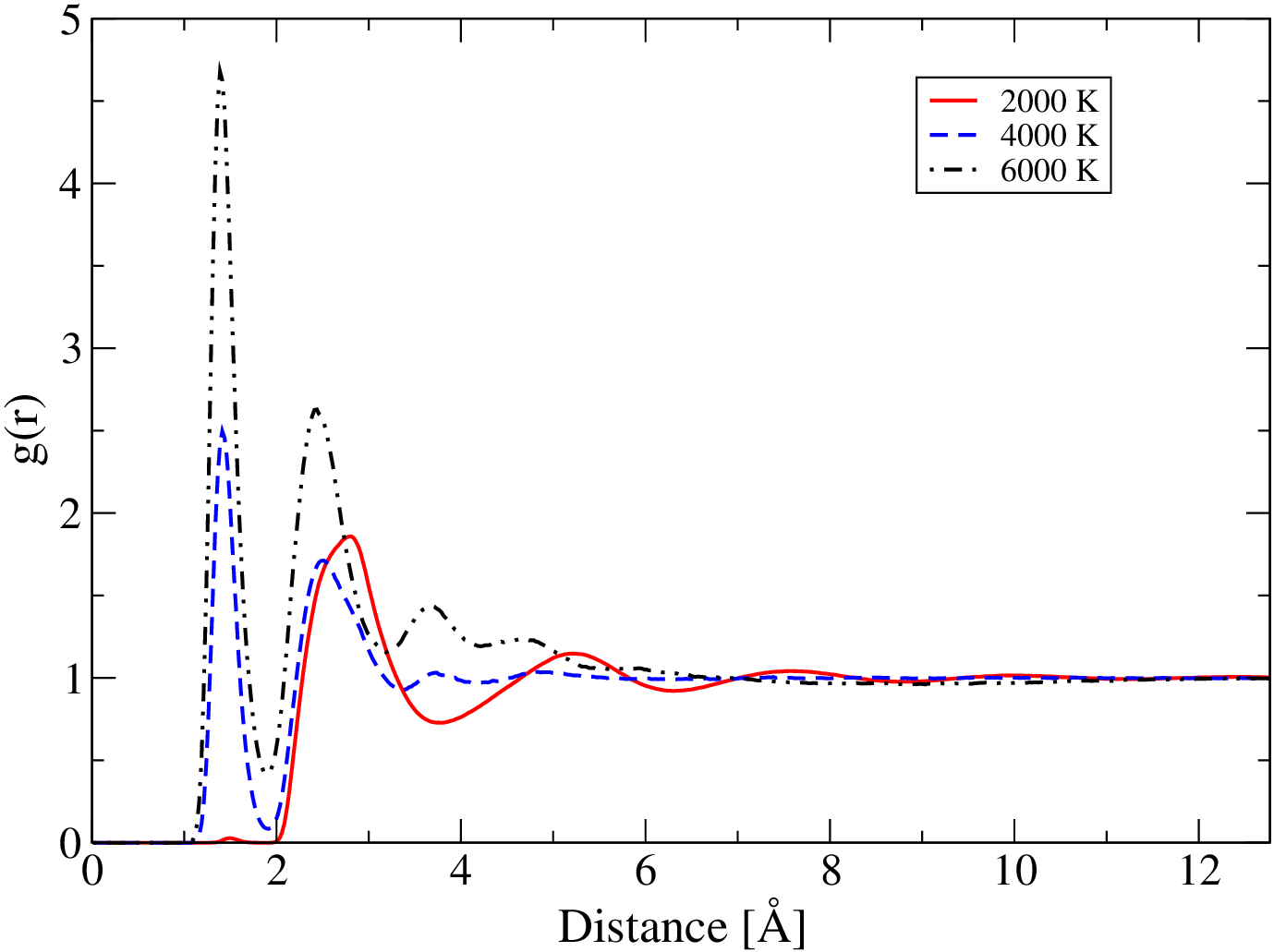}
\caption{(Color online) Comparison of the C-C PCF at 2000 K (solid red line), 4000 K (dashed blue line) and 6000 K (dot-dashed black line).} 

\label{fig:rdfCC}
\end{figure}

\begin{figure}
\includegraphics[width=0.475\textwidth]{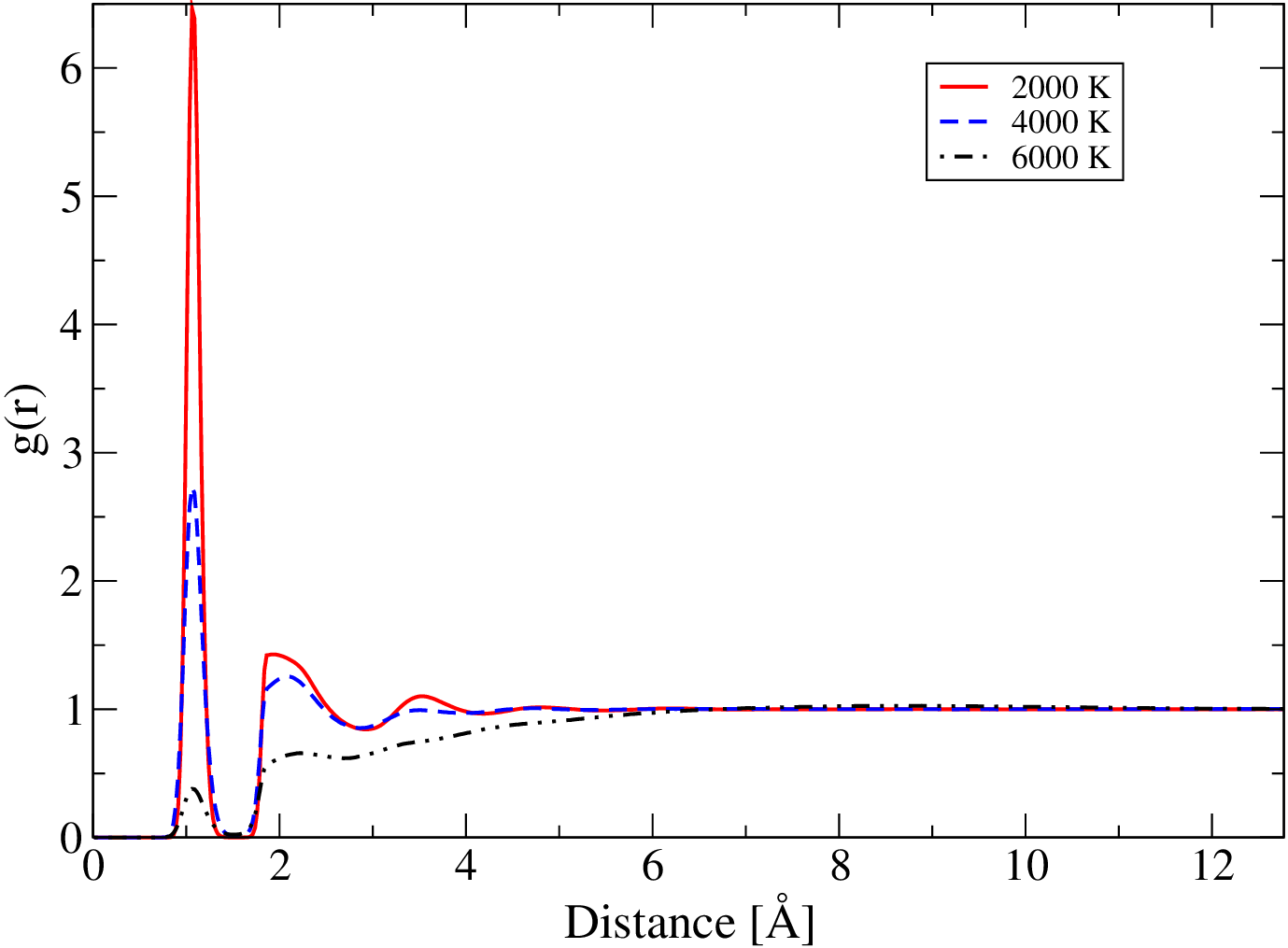}
\caption{(Color online) The C-H PCF at 2000 K (solid red line), 4000 K (dashed blue line) and 6000 K (dot-dashed black line).} 
\label{fig:rdfCH}
\end{figure}

\begin{figure}
\includegraphics[width=0.475\textwidth]{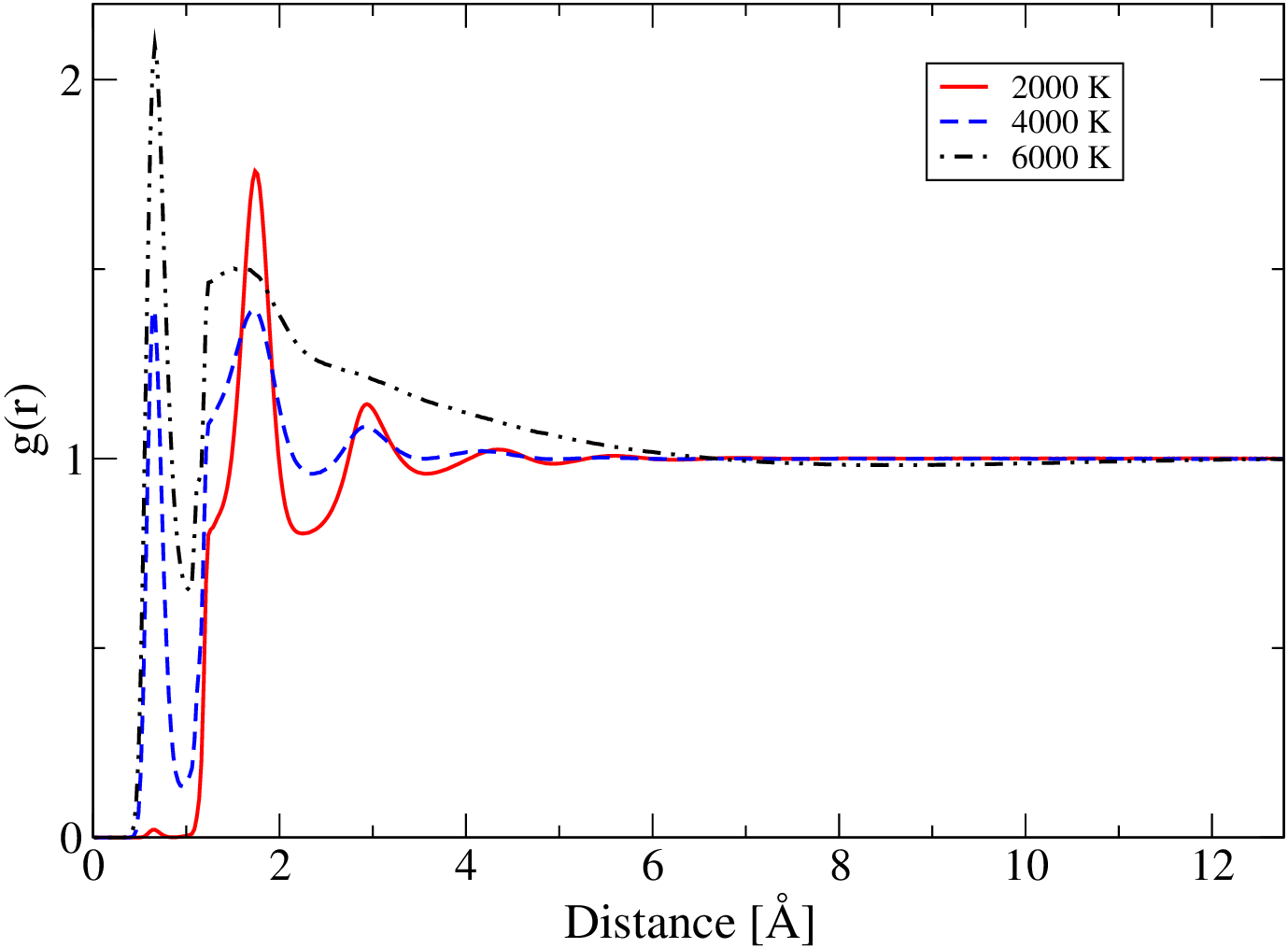}
\caption{(Color online) The H-H PCF at 2000 K (solid red line), 4000 K (dashed blue line) and 6000 K (dot-dashed black line).} 
\label{fig:rdfHH}
\end{figure}

The partial pair correlation functions (PCF) of our simulations are shown in Figs.~\ref{fig:rdfCC}-\ref{fig:rdfHH}. As can be seen in Fig.~\ref{fig:rdfCC}, as well as Fig.~\ref{fig:rdfHH}, at $T=2000$~K essentially no covalent C-C and H-H bonds are present. The only remaining significant peaks at $2.81\,$~\r{A} and $1.74\,$~\r{A} represent the average C-C and H-H distances between two adjacent CH$_4$ molecules, respectively. The insignificant peaks at $1{.}47\,$ \r{A} in Fig.~\ref{fig:rdfCC} and $0{.}75\,$ \r{A} in Fig.~\ref{fig:rdfHH} do not point to an onset of dissociation, as proposed by experiment \cite{ross, benedetti}, but are rather due to fleetingly broken C-H bonds caused by finite temperature. Consequently, the sharp intramolecular peak in Fig.~\ref{fig:rdfCH} at around $1.075\,$~\r{A} can be ascribed to covalent C-H bonds. The corresponding coordination numbers, as obtained by integrating the associated PCF's up to their first minima, are shown in Table~\ref{coord}. In the case of C-H the partial coordination number is $3.984$, which indicates that the liquid at $T=2000$~K, except for single fleetingly broken C-H bonds, is nearly exclusively made up of undissociated CH$_4$ molecules. This view is consistent with other AIMD studies \cite{ancil, kress1, spanu}, but at variance to theoretical ground-state calculations \cite{ree1}, as well as experimental measurements \cite{nellis2, benedetti, ross}, no signature for dissociation have been found.

For the most relevant case at $T=4000$~K  and $P \approx 100$~GPa the situation is much different and evidences for dissociation can indeed be observed. As can be seen in Fig.~\ref{fig:rdfCC} covalent C-C bonds are appearing just like covalently bonded H$_2$ dimers, as shown in Fig.~\ref{fig:rdfHH}. As a consequence, the height of the intramolecular C-H peak in Fig.~\ref{fig:rdfCH} is much reduced, though still existing. From Table~\ref{coord} it can be deduced that nearly half of the covalent C-H bonds are broken, which indicates that methane does dissociate only partially to form hydrocarbon chains with mainly two and three carbon atoms, as well as H$_2$. 
More precisely, the CH$_4$ molecules dissociate and recombine to form C$_2$H$_6$ and to a smaller extend C$_3$H$_8$, which is in agreement with previous AIMD studies \cite{ancil, kress1, spanu}. However, we find no sustained signature for the presence of C$_2$H$_2$, which has been detected in the atmosphere of Neptune \cite{conrath}. But, we do find seeds of somewhat longer sp$^2$-bonded chains and ring-like carbon structures, but definitely no signature of sp$^3$ carbon bonds, i.e. no diamond-like carbon. This is consistent with the computed vibrational density of states of Spanu et al. \cite{spanu}, who report a noticeable feature at 1600~cm$^{-1}$ that can be attributed to threefold coordinated carbon atoms in graphite-like configurations. 
That is to say, that on the one hand our calculations are in agreement with experiment by implying that in the middle ice layer CH$_4$ molecules itself are not present, but merely its dissociated constituents, which confirms that the interior chemistry of Uranus and Neptune is more complex than previously assumed. On the other hand, our results differ in the sense that at $T=4000$~K we do not find any evidence for diamond-like carbon, as reported by the very same experiments \cite{nellis2, benedetti, ross}. 

Even deeper within the planet at even higher temperature of $T=6000$~K, the remaining CH$_4$ have fully dissociated as indicated by the vanishing intramolecular C-H peak in Fig.~\ref{fig:rdfCH}. On the other hand, the first peak in Fig.~\ref{fig:rdfCC}, which is due to C-C bonds, as well as the covalent H-H peak in Fig.~\ref{fig:rdfHH} are even more pronounced as is the case for $T=4000$~K. As shown in Table~\ref{coord} the partial C-H coordination number is rather small, which entails that contrary to $T=4000$~K hydrocarbon chains are no longer present, but have completely dehydrogenated into polymeric carbon and hydrogen. As before, no evidence for sp$^3$-bonded diamond could be found, which indicates that even higher pressures are required to condense carbon into diamond.

\begin{table}
  \caption{Partial coordination numbers, as obtained by integrating the associated PCFs up to their first minima, for all investigated temperatures.}
  \label{coord}
  \begin{center}
    \begin{tabular}{|c|c|c|c|}
      \hline\hline
      Temperature & \quad C-C \quad & \quad C-H \quad & \quad H-H \quad \\
      \hline
      $2000$~K & \quad $0.080$ \quad & \quad $3.984$ \quad & \quad $0.004$ \quad \\
      $4000$~K & \quad $1.244$ \quad & \quad $2.298$ \quad & \quad $0.408$ \quad \\
      $6000$~K & \quad $2.556$ \quad & \quad $0.417$ \quad & \quad $1.115$ \quad \\
      \hline\hline
    \end{tabular}
  \end{center}
\end{table}

However, in contrast to previous AIMD simulations \cite{ancil, kress1, spanu} we find for the first time that at $T = 6000$~K hydrogen is no longer solely molecular, but a noticeable fraction of monoatomic hydrogen can be identified. This immediately suggest that at $T=6000$~K the present hydrogen molecules are on the verge of a liquid-liquid phase transition into an atomic fluid phase, which is in agreement with recent calculations \cite{tamblyn, morales}. Due to the fact that dissociation is believed to be accompanied with the metallization \cite{scandolo, azadi}, our prediction of liquid atomic hydrogen may lead to an explanation for the large magnetic fields of planets such as Uranus and Neptune through a dynamo-like mechanism by electrical currents in the liquid metallic regions of their interiors \cite{ness1, *ness2}. Since we find that liquid CH$_4$, where it is stable at $T=2000$~K, is a wide band-gap insulator, it can only fully dehydrogenated contribute to the magnetic field in the form of metallic hydrogen, which indeed has been established experimentally at rather similar conditions by shock-compression experiments \cite{nellis3, nellis4}. 
Moreover, the motion of charged particles trapped in such magnetic fields causes the generation of radio waves. In fact, planetary radio experiments aboard the Voyager~II flyby mission detected a wide variety of of radio emissions for both planets \cite{warwick1, *warwick2}. 

If large enough, amorphous or crystalline carbon clusters precipitate and sink towards the planetary center as sediment via gravitational settling. The corresponding release of energy has been estimated to be a substantial fraction of the internal heat production, which would explain for instance why Neptune radiates more than twice the energy it receives from the sun \cite{benedetti}. Furthermore, it is also likely the cause for the externally observed high luminosity and could even contribute to the convective motions of its fluid interior of Neptune. The reason, why for Uranus no such internal heat flow mechanism could be established is still unknown \cite{hubbardNeptune}. Nevertheless, the similarity of the internal structures of these two planets suggests that the suppressed convection of Uranus may be a consequence of its closer proximity to the sun. In contrast, saturated hydrocarbons such as C$_2$H$_6$ and H$_2$, being the products of the above ascertained decomposition of CH$_4$ at $T=4000$~K, do not precipitate and instead rather rise to join the atmosphere. 
As a consequence, this process could be responsible for the anomalous abundance of H$_2$ in the atmospheres of both planets, and in the case of Neptune may also account for the observed wealth of atmospheric C$_2$H$_6$, where it might be brought up from the deep interior by the just elucidated convection process. Therefore, the present results imply that deep chemical processes such as phase transformations at extreme temperatures and pressures must be considered to model the interiors of giant gas planets more realistically.

Even though, our calculations provide a consistent picture of the deep chemistry of Neptune and Uranus, 
the remaining question is why no diamond formation could be observed, whereas experimentally it is reported to occur from $P=20$~GPa and $T=2000$~K on. 
Due to the fact that liquid methane is optically transparent and can not simply be heated by a laser beam, it is therefore common practice to include a noble metal absorber within laser-heated diamond anvil cell experiments. Spanu et al. reported that without a metallic absorber no formation of complex hydrocarbons and H$_2$ at $T=2000$~K could be determined, which not only agrees with the findings of the present work but also indicates that liquid CH$_4$ resides in a metastable state. On the contrary, at the presence of a nobel metal, liquid CH$_4$ readily dissociates \cite{spanu}. 

We conclude by noting that another possibility to explain the discrepancy between theory and experiment may be the existence of a homogenous nucleation mechanism, similar to the one recently proposed by Khaliullin et al. for the direct graphite-to-diamond transition \cite{khaliullin1, *khaliullin2, ghiringhelli}. 

\begin{acknowledgments}
The authors acknowledge financial support from the Graduate School of Excellence MAINZ and the IDEE project of the Carl Zeiss Foundation. We are grateful to Manuel D\"omer for critical reading the manuscript.  
\end{acknowledgments}

\bibliography{paper_m_tdk}

\end{document}